

\font\eighteenrm = cmr10 scaled\magstep3
\font\eighteeni = cmmi10 scaled\magstep3
\font\eighteensy = cmsy10 scaled\magstep3
\font\eighteenit = cmti10 scaled\magstep3
\font\eighteenb = cmbx10 scaled\magstep3

\font\twelverm = cmr12

\font\twelvei = cmmi12
\font\twelveit = cmti12
\font\twelveb = cmbx12
\font\twelvesy = cmsy10 scaled\magstep1
\font\twelves = cmsl12

\font\tenrm = cmr10
\font\tenit = cmti10
\font\teni = cmmi10                  
\font\tensy = cmsy10                
\font\tens = cmsl10                  
\font\tenb = cmbx10                  

\font\teniu = cmu10
\font\ninerm = cmr9
\font\ninesy = cmsy9
\font\nineb = cmbx9

\font\eightrm = cmr8
\font\eighti = cmmi8
\font\eightsy = cmsy8

\font\sixrm = cmr6
\font\sixi  = cmmi6
\font\sixsy = cmsy6

\font\fivesy = cmsy5

\def\tenpoint{\def\rm{\fam0\tenrm}
\textfont0=\tenrm \scriptfont0=\eightrm \scriptscriptfont0=\sixrm
\textfont1=\teni \scriptfont1=\eighti \scriptscriptfont1=\sixi
\textfont2=\tensy \scriptfont2=\eightsy \scriptscriptfont2=\sixsy
\textfont3=\tenex \scriptfont3=\tenex \scriptscriptfont3=\tenex
\def\sy{\fam4\tensy}%
\textfont4=\tensy%
\def\sl{\fam5\tens}%
\textfont5=\tens%
\def\bf{\fam6\tenb}%
\textfont6=\tenb%
\def\it{\fam7\tenit}%
\textfont7=\tenit%
\def\prfnt{\fam8\ninesy}%
\textfont8=\ninesy%
\def\hbfnt{\fam9\fivesy}%
\textfont9=\fivesy%
\textfont11=\sixrm \scriptfont11=\sixrm \scriptscriptfont11=\sixrm%
\baselineskip 12pt                                                
\lineskip 1pt
\parskip 5pt plus 1pt
\abovedisplayskip 12pt plus 3pt minus 9pt
\belowdisplayshortskip 7pt plus 3pt minus 4pt \tenrm}
\def\prfnt{\ninesy }
\def\hbfnt{\fivesy }

\def\bigfnt{\twelverm}

\def\eighteenpoint{\def\rm{\fam0\eighteenrm}%
\textfont0=\eighteenrm \scriptfont0=\twelverm \scriptscriptfont0=\eightrm
\textfont1=\eighteeni \scriptfont1=\twelvei \scriptscriptfont1=\eighti
\textfont2=\eighteensy \scriptfont2=\twelvesy \scriptscriptfont2=\eightsy
\textfont3=\tenex \scriptfont3=\tenex \scriptscriptfont3=\tenex
\def\sy{\fam4\eighteensy}%
\textfont4=\eighteensy%
\def\bf{\fam6\eighteenb}%
\textfont6=\eighteenb%
\def\it{\fam7\eighteenit}%
\textfont7=\eighteenit%
\baselineskip 21pt
\lineskip 1pt
\parskip 5pt plus 1pt
\abovedisplayskip 15pt plus 5pt minus 10pt
\belowdisplayskip 15pt plus 5pt minus 10pt
\abovedisplayshortskip 13pt plus 8pt
\belowdisplayshortskip 10pt plus 5pt minus 5pt
\eighteenrm}
\def\twelvepoint{\def\rm{\fam0\twelverm}%
\textfont0=\twelverm \scriptfont0=\tenrm \scriptscriptfont0=\eightrm
\textfont1=\twelvei \scriptfont1=\teni \scriptscriptfont1=\eighti
\textfont2=\twelvesy \scriptfont2=\tensy \scriptscriptfont2=\eightsy
\textfont3=\tenex \scriptfont3=\tenex \scriptscriptfont3=\tenex
\def\sy{\fam4\twelvesy}%
\textfont4=\twelvesy%
\def\sl{\fam5\twelves}%
\textfont5=\twelves%
\def\bf{\fam6\twelveb}%
\textfont6=\twelveb%
\def\it{\fam7\twelveit}%
\textfont7=\twelveit%
\def\prfnt{\fam8\ninesy}%
\textfont8=\ninesy%
\def\hbfnt{\fam9\fivesy}%
\textfont9=\fivesy%
\def\paperfont{\fam10\twelverm}%
\def\dotfont{\fam11\sixrm}%
\textfont11=\sixrm \scriptfont11=\sixrm \scriptscriptfont11=\sixrm%
\baselineskip 15pt
\lineskip 1pt
\parskip 5pt plus 1pt
\abovedisplayskip 15pt plus 5pt minus 10pt
\belowdisplayskip 15pt plus 5pt minus 10pt
\abovedisplayshortskip 13pt plus 8pt
\belowdisplayshortskip 10pt plus 5pt minus 5pt \twelverm}

\tenpoint
\hsize 5.9truein
\vsize 9.5truein
\global \topskip .7truein
\newcount\chapnum
\chapnum = 0
\newcount\sectnum
\sectnum = 0
\newcount\subsecnum            
\subsecnum = 0        
\newcount\eqnum
\newcount\chapnum
\chapnum = 0
\newcount\sectnum
\sectnum = 0
\newcount\subsecnum            
\subsecnum = 0        
\newcount\eqnum
\eqnum = 0
\newcount\refnum
\refnum = 0
\newcount\tabnum
\tabnum = 0
\newcount\fignum
\fignum = 0
\newcount\footnum
\footnum = 1
\newcount\pointnum
\pointnum = 0
\newcount\subpointnum
\subpointnum = 96
\newcount\subsubpointnum
\subsubpointnum = -1
\newcount\letnum
\letnum = 0
\newbox\referens
\newbox\figures
\newbox\tables
\newbox\tempa
\newbox\tempb
\newbox\tempc
\newbox\tempd
\newbox\tempe
\hbadness=10000
\newbox\refsize
\setbox\refsize\hbox to\hsize{ }
\hbadness=1000
\newskip\refbetweenskip
\refbetweenskip = 5pt
\def\ctrline#1{\line{\hss#1\hss}}
\def\rjustline#1{\line{\hss#1}}
\def\ljustline#1{\line{#1\hss}}

\def\ctr#1{\hfill{#1}\hfill}

\def\spose#1{\hbox to 0pt{#1\hss}}
\def\chskipt{\vskip .125in plus 0pt minus 0pt }              
\def\chskipl{\vskip .7in plus 18pt minus 10pt}               
\def\secskipt{\penalty-500\vskip 24pt plus 2pt minus 2pt}    
\def\secskipl{\vskip 3.5pt plus 1pt }
\def\subsecskip{\penalty-500\vskip 6pt plus 2pt minus 2pt }
\def\unchskip{\vskip -.7in }                                 
\def\conskip{\vskip 14pt }
\newif\ifoddeven
\gdef\oneside{\oddevenfalse}

\oneside
\newif\ifnonumpageone

\gdef\nonumberfirst{\nonumpageonetrue}
\nonumberfirst
\output{\ifoddeven\leftright\else\samemarg\fi
        \ifnonumpageone\checkpage\else\empty\fi
        \plainoutput}
\def\leftright{\ifodd\count0{\global\hoffset=\oddmargin}
                       \else{\global\hoffset=\evenmargin}\fi}
\def\samemarg{\global\hoffset=\oddmargin}
\gdef\oddmargin{.25truein}
\gdef\evenmargin{0truein}
\def\checkpage{\ifnum\count0=1\nopagenumbers\else\empty\fi}
\footline={{\pagefont\hss--\hquad\folio\hquad--\hss}}
\def\pagefont{\teniu}
\setbox\referens\vbox{\ctrline{\bf References }\chskipt }
\setbox\figures\vbox{\ctrline{\bf Figure Captions }\chskipt }
\setbox\tables\vbox{\ctrline{\bf Table Captions }\chskipt }

\def\title#1{\ctrline {\titfnt #1} }
\def\titfnt{\eighteenpoint}
\def\author#1{\ctrline{\autfnt #1}\par}
\def\autfnt{\bigfnt}

\def\abstract{
\ctrline{\bf ABSTRACT}\chskipt}

\def\reset{\global\sectnum = 0 \global\eqnum = 0
     \global\subsecnum = 0}
\def\chap#1{\global\advance\chapnum by 1 \reset 
\endpage
\chskipt
\ifodd\count0{
\rightline{{\chapnumfont Chapter \the\chapnum}}
\medskip
\rightline{{\chapfont #1}}}
\else{
\leftline{{\chapnumfont Chapter \the\chapnum}}
\medskip
\ljustline{{\chapfont #1}}}\fi
\penalty 100000 \chskipl  \penalty 100000
{\let\number=0\edef\next{
\write2{\bigskip\noindent
  \tofcfont Chapter \the\chapnum.{ }#1
  \leadtofc\number\count0\smallskip}}
\next}}
\def\titcon#1{\unchskip
\ifodd\count0{
\rightline{{\chapfont #1}}}
\else{
\leftline{{\chapfont #1}}}\fi
\penalty 100000 \chskipl \penalty 100000 }
\def\chapnumfont{\tenit}
\def\chapfont{\eighteenpoint}
\def\chapter#1{\chap{#1}}
\def\sect#1{\global\advance\sectnum by 1 \global\subsecnum = 0
\secskipt
\ifnum\chapnum=0{{\sectfont\par\noindent
\secsign\the\sectnum{ }{ }#1}\par
         {\let\number=0\edef\next{
         \write2{\vskip 0pt\noindent\hangindent 30pt
         \tofcfont\hbox to 30pt{\hfill\the\sectnum\quad}\unskip#1
         \leadtofc\number\count0}}
         \next}}
\else{{\sectfont\par\noindent
\secsign\the\chapnum .\the\sectnum{ }{ }#1}\par
      {\let\number=0\edef\next{
       \write2{\vskip 0pt\noindent\hangindent 30pt
       \tofcfont\hbox to 30pt{\hfill\the\chapnum
              .\the\sectnum\quad}\unskip#1
         \leadtofc\number\count0}}
       \next}}\fi
       \nobreak\medskip\nobreak}
\def\sectfont{\twelvepoint}
\def\secsign{\S}

\def\subsect#1{\global\advance\subsecnum by 1 \secskipt
\noindent
\ifnum\chapnum=0{
     {\subsecfont\the\sectnum .\the\subsecnum{ }{ }#1}
      \nobreak\par
     {\let\number=0\edef\next{
     \write2{\vskip 0pt\noindent\hangindent 58pt\tofcfont
     \hbox to 60pt{\hfill\the\sectnum
           .\the\subsecnum\quad}\unskip#1
     \leadtofc\number\count0}}\next}}
\else{{
     \subsecfont\the\chapnum .\the\sectnum
        .\the\subsecnum{ }{ }#1}\nobreak\par
     {\let\number=0\edef\next{
     \write2{\vskip 0pt\noindent\hangindent 58pt\tofcfont
     \hbox to 60pt{\hfill\the\chapnum.\the\sectnum
           .\the\subsecnum\quad}\unskip#1
     \leadtofc\number\count0}}\next}}\fi
     \nobreak\medskip\nobreak}

\def\subsecfont{\twelvepoint}        
\immediate\openout 2 tofc
\def\tableofcontents#1{\endpage
\count0=#1
\chskipt
\ifodd0{
\rjustline{{\chapfont Contents}}}
\else{
\ljustline{{\chapfont Contents}}}\fi
\chskipl
\rjustline{{\tofcfont Page}}
\bigskip
\immediate\closeout 2
\input tofc
\endpage}

\def\tofcfont{\ninerm}
\def\leadtofc{\leaders\hbox to 8pt{\hfill.\hfill}\hfill}
\immediate\openout 4 refc
\def\refbegin#1#2{\unskip\global\advance\refnum by1
\xdef\rnum{\the\refnum}
\xdef#1{\the\refnum}
\xdef\rtemp{$^{\rnum}$}
\unskip
\immediate\write4{\vskip 5pt\par\noindent\tofcfont
  \hangindent .11\wd\refsize \hbox to .11\wd\refsize{\hfill
  \the\refnum . \quad } \unskip}\unskip
  \immediate\write4{#2}\unskip}
\def\refend{\nobreak\rtemp\unskip}
\def\ref#1{\refbegin{\?}{#1}}

\def\REF#1#2{\refbegin{#1}{#2}}
\def\refsbegin#1#2{\unskip\global\advance\refnum by 1
\xdef\refb{\the\refnum}
\xdef#1{\the\refnum}
\xdef\rrnum{\the\refnum}
\unskip
\immediate\write4{\vskip 5pt\par\noindent\tofcfont
  \hangindent .11\wd\refsize \hbox to .11\wd\refsize{\hfill
  \the\refnum . \quad } \unskip}\unskip
  \immediate\write4{#2}\unskip}
\def\REFSCON#1#2{\unskip \global\advance\refnum by 1
\xdef#1{\the\refnum}
\xdef\rrrnum{\the\refnum}
\unskip
\immediate\write4{\vskip 5pt\par\noindent\tofcfont
  \hangindent .11\wd\refsize \hbox to .11\wd\refsize{\hfill
  \the\refnum . \quad } \unskip}\unskip
  \immediate\write4{#2}\unskip}

\def\refsend{\nobreak$^{\refb-\the\refnum}$\unskip}

\def\REFS#1#2{\refsbegin{#1}{#2} }
\def\foot#1{\footnote{$^{\the\footnum}$}{#1}
  \global\advance\footnum by 1}

\def\figure#1#2{\global\advance\fignum by 1
\xdef#1{\the\fignum }
\ctrline{\Figure . #2}\par\conskip
{\let\number=0\edef\next{
\write3{\par\noindent\tofcfont
  \hangindent .11\wd\refsize \hbox to .11\wd\refsize{\hfill
  \the\fignum . \quad } \unskip}
\write3{#2\leadtofc\number\count0\par}}
\next}}
\def\figurs#1#2#3{\global\advance\fignum by 1
\xdef#1{\the\fignum }
\ctrline{\Figure . \it #2}
\ctrline{\it #3}\par\conskip
{\let\number=0\edef\next{
\write3{\par\noindent\tofcfont
  \hangindent .11\wd\refsize \hbox to .11\wd\refsize{\hfill
  \the\fignum . \quad}\unskip}
\write3{#2 #3\leadtofc\number\count0\par}}
\next}}

\def\figcon{\ctrline{{\it Figure  \the\fignum} -- cont'd}\par\conskip}
\def\Figure{{\it Figure  \the\fignum}}
\immediate\openout 3 figc

\def\table#1#2{\global\advance\tabnum by 1
\xdef#1{\the\tabnum }
\ctrline{\Table . #2}\par\conskip
{\let\number=0\edef\next{
\write5{\par\noindent\tofcfont
  \hangindent .11\wd\refsize \hbox to .11\wd\refsize{\hfill
  \the\tabnum . \quad } \unskip}
\write5{#2\leadtofc\number\count0\par}}
\next}}
\def\tabls#1#2#3{\global\advance\tabnum by 1
\xdef#1{\the\tabnum }
\ctrline{\Table . #2}
\ctrline{\it #3}\par\conskip
{\let\number=0\edef\next{
\write5{\par\noindent\tofcfont
  \hangindent .11\wd\refsize \hbox to .11\wd\refsize{\hfill
  \the\tabnum . \quad}\unskip}
\write5{#2 #3\leadtofc\number\count0\par}}
\next}}

\def\Table{\it Table  \the\tabnum}
\immediate\openout 5 tabc

\def\eqname#1{\global\advance\eqnum by 1
\ifnum\chapnum=0{
   \xdef#1{ (\the\eqnum ) }(\the\eqnum )  }
\else{
   \xdef#1{ (\the\chapnum .\the\eqnum ) }
            (\the\chapnum .\the\eqnum ) }\fi}
\def\enum{\global\advance\eqnum by 1
  \ifnum\chapnum=0{ (\the\eqnum )  }
  \else{(\the\chapnum .\the\eqnum ) }\fi}

\def\eqn#1{\eqno \eqname{#1} }
\def\eqnameap#1{\global\advance\eqnum by 1
   \xdef#1{ (\copy\appbox .\the\eqnum ) }
            (\copy\appbox .\the\eqnum ) }
\def\enumap{\global\advance\eqnum by 1
  (\copy\appbox .\the\eqnum ) }

\def\item#1{\par\noindent\hangindent .08\wd\refsize
\hbox to .08\wd\refsize{\hfill #1\quad}\unskip}
\def\sitem#1{\par \noindent\hangindent .13\wd\refsize
\hbox to .13\wd\refsize{\hfill #1\quad}\unskip}
\def\ssitem#1{\par\noindent\hangindent .195\wd\refsize
\hbox to .195\wd\refsize{\hfill #1\quad}\unskip}

\def\point{\par \global\advance\pointnum by 1
\noindent\hangindent .08\wd\refsize \hbox to .08\wd\refsize{\hfill
\the\pointnum .\quad}\unskip}

\def\spoint{\par \global\advance\subpointnum by 1
\noindent\hangindent .13\wd\refsize
\hbox to .13\wd\refsize{\hfill
(\char\the\subpointnum )\quad}\unskip}

\def\sspoint{\par \global\advance\subsubpointnum by 1
\noindent\hangindent .195\wd\refsize
\hbox to .195\wd\refsize {\hfill\hbox to 20pt{
(\romannumeral\subsubpointnum\hfill)}\quad}\unskip}

\def\bye{\endpage\end}              

\def\bspace#1{\hbox to -#1{}}
\newbox\appbox
\def\appendix#1{\endpage\reset
\setbox\appbox\hbox{#1}
\chskipt \ctrline {\bf APPENDIX #1 }\penalty 10000
\chskipl \penalty 10000
\write2{\bigskip\noindent
 {\tofcfont Appendix #1\leadtofc\number\count0\par\smallskip}}}

\def\mat#1#2{\if 2#1 {\left( \  \vcenter{\halign{$\ctr{## }$ \quad
& $\ctr{## }$\cr #2}} \  \right) } \else{ }\fi
\if 3#1 {\left( \  \vcenter{\halign{
$\ctr{## }$ \quad & $\ctr{## }$ \quad
& $\ctr{## }$\cr #2}} \  \right) } \else{ }\fi
\if 4#1 {\left( \  \vcenter{\halign{$\ctr{## }$ \quad &
$\ctr{## }$ \quad & $\ctr{## }$ \quad
& $\ctr{## }$\cr #2}} \  \right) } \else{ }\fi
\if 5#1 {\left( \  \vcenter{\halign{$\ctr{## }$ \quad
& $\ctr{## }$ \quad & $\ctr{## }$ \quad
& $\ctr{## }$ \quad & $\ctr{## }$\cr #2}} \  \right)} \else{ }\fi
\if 6#1 {\left( \  \vcenter{\halign{$\ctr{## }$ \quad
& $\ctr{## }$ \quad & $\ctr{## }$ \quad & $\ctr{## }$ \quad
& $\ctr{## }$ \quad & $\ctr{## }$\cr #2}} \  \right)} \else{ }\fi }

\def\endpage{\par \vfill \eject}
\def\physrev{\baselineskip 24pt
\lineskip 1pt
\parskip 1pt plus 1pt
\abovedisplayskip 15pt plus 7pt minus 13.33pt
\belowdisplayskip 15pt plus 7pt minus 13.33pt
\abovedisplayshortskip 14pt plus 11pt
\belowdisplayshortskip 9pt plus 7pt minus 7pt
\def\chskipt{\vskip 24pt }
\def\chskipl{\vskip 6.5pt }
\def\secskipt{\vskip 7pt plus 3pt minus 1.33pt }
\def\secskipl{\vskip 3.5pt plus 2pt }
\def\subsecskip{\vskip 7pt plus 2pt minus 2pt }
\def\unchskip{\vskip -6.5pt }
\def\conskip{\vskip 24pt }
\refbetweenskip = \the\baselineskip
\multskip\refbetweenskip by 5
\divskip\refbetweenskip by 10
\twelverm }
\def\bk{\hfil\break}         

\def\To{\par\noindent\hangindent .18\wd\refsize
\hbox to .18\wd\refsize {To: \hfill \qquad }}
\def\from{\par\noindent\hangindent .18\wd\refsize
\hbox to .18\wd\refsize {From: \hfill \qquad }}
\def\topic{\par\noindent\hangindent .18\wd\refsize
\hbox to .18\wd\refsize{Topic: \hfill \qquad }}

\def\startpage#1 {\count0 = #1}
\def\startchapter#1{\chapnum = #1 \advance\chapnum by -1}

\def\startfig#1{\fignum = #1 \advance\fignum by -1}
\def\starttab#1{\tabnum = #1 \advance\tabnum by -1}

\newbox\xa\newbox\xb
\def\boxit#1{ \setbox\xa\vbox {\vskip \boxitsep
\hbox{\hskip \boxitsep #1\hskip \boxitsep }\vskip \boxitsep }
\setbox\xb\hbox{\vrule \copy\xa \vrule}
\vbox{\hrule width 1\wd\xb \copy\xb \hrule width 1\wd\xb }}
\def\fboxit#1#2{ \setbox\xa\vbox {\vskip \boxitsep
\hbox{\hskip \boxitsep #2\hskip \boxitsep }\vskip \boxitsep }
\setbox\xb\hbox{\vrule width #1pt \copy\xa \vrule width #1pt}
\vbox{\hrule height #1pt width 1\wd\xb
\copy\xb \hrule height #1pt width 1\wd\xb }}
\def\reboxit#1#2#3{ \setbox\xa\vbox{\vskip \boxitsep
\hbox{\hskip \boxitsep #3\hskip \boxitsep }\vskip \boxitsep }
\setbox\xb\hbox{\vrule width #1pt\bspace{#2}
\copy\xa \vrule width #1pt}
\vbox{\hrule height #1pt width 1\wd\xb
\copy\xb \hrule height #1pt width 1\wd\xb}}
\def\boxitsep{4pt}

\newdimen\offdimen
\def\offset#1#2{\offdimen #1
   \noindent \hangindent \offdimen
   \hbox to \offdimen{#2\hfil}\ignorespaces}
\newdimen\defnamespace   
\defnamespace=2in        
\def\definition#1#2{     
    \def\itema{\par\hang\textindent}
    {\advance\parindent by \defnamespace
     \advance\defnamespace by -.5em
     \itema{\hbox to \defnamespace{#1\hfil}}#2\par}}
\def\TEX{\hbox{T\hskip-2pt\lower1.94pt\hbox{E}\hskip-2pt X}}
\def\wyz{\hbox{WI\hskip-1pt\lower.9pt\hbox{Z\hskip-1.85pt
\raise1.7pt\hbox{Z}}LE}}
\def\\{$\backslash $}

\def\underwiggle#1{\mathop{\vtop{\ialign{##\crcr
    $\hfil\displaystyle{#1}\hfil$\crcr\noalign{\kern2pt\nointerlineskip}
    $\scriptscriptstyle\sim$\crcr\noalign{\kern2pt}}}}\limits}
\def\({[}
\def\){]}

\def\hquad{\hskip.5em{}}
\mathchardef\app"3218

\def\linebreak{\break}
\mathchardef\oprod="220A
\mathchardef\inter="225C
\mathchardef\union="225B

\mathchardef\relv="326A
\mathchardef\leftv="326A
\mathchardef\rightv="326A
\mathchardef\relvv"326B
\mathchardef\leftvv"326B
\mathchardef\rightvv"326B
\mathchardef\Zscr"25A
\mathchardef\Yscr"259
\mathchardef\Xscr"258
\mathchardef\Wscr"257
\mathchardef\Vscr"256
\mathchardef\Uscr"255
\mathchardef\Tscr"254
\mathchardef\Sscr"253
\mathchardef\Rscr"252
\mathchardef\Qscr"251
\mathchardef\Pscr"250
\mathchardef\Oscr"24F
\mathchardef\Nscr"24E
\mathchardef\Mscr"24D
\mathchardef\Lscr"24C
\mathchardef\Kscr"24B
\mathchardef\Jscr"24A
\mathchardef\Iscr"249
\mathchardef\Hscr"248
\mathchardef\Gscr"247
\mathchardef\Fscr"246
\mathchardef\Escr"245
\mathchardef\Dscr"244
\mathchardef\Cscr"243
\mathchardef\Bscr"242
\mathchardef\Ascr"241
\mathchardef\lscr"160

\immediate\openout 2 tofc
\immediate\openout 3 figc
\immediate\openout 4 refc
\immediate\openout 5 tabc
\tolerance 4000
\def\refbegin#1#2{\unskip\global\advance\refnum by1
\xdef\rnum{\the\refnum}
\xdef#1{\the\refnum}
\xdef\rtemp{[\rnum]}
\unskip
\immediate\write4{\vskip 5pt\par\noindent\tofcfont
  \hangindent .11\wd\refsize \hbox to .11\wd\refsize{\hfill
  \the\refnum . \quad } \unskip}\unskip
  \immediate\write4{#2}\unskip}
\def\refsend{\nobreak[\refb-\the\refnum]\unskip}
\font\mss=cmsy10
\def\one{1}
\def\rar{\rightarrow}
\def\_#1{_{\scriptscriptstyle #1}}
\def\^#1{^{\scriptscriptstyle #1}}
\def\vt{virial theorem}
\def\vrt#1{\vec r\_{#1}(t)}
\def\eom{equation of motion}
\def\eoms{equations of motion}
\def\dof{degree of freedom}
\def\dofs{degrees of freedom}

\def\deriv#1#2{{d#1\over d#2}}
\def\oot{{1\over 2}}
\def\pdline#1#2{\partial#1/\partial#2}
\def\av#1{\langle#1\rangle}
\def\avlar#1{\big\langle#1\big\rangle}
\def\A{A}
\def\FF{F}
\def\EPS{\epsilon}
\def\M{M}
\def\A{{\hbox{\mss\char'101}}}
\def\FF{{\hbox{\mss\char'106}}}
\def\Msmall{{{\scriptscriptstyle\mss\char'115}}}
\def\M{{\hbox{\mss\char'115}}}
\def\EPS{{\hbox{\mss\char'105}}}
\def\AM{\A\_{\Msmall}}
\def\nineb{\bf}
\def\ninerm{}
\def\({[}
\def\){]}
\def\dtr{d\^3r}

\def\dfxg{g\^{1/2}d\^{4}x}

\def\grad{\vec\nabla}

\def\b{\beta}
\def\d{\delta}
\def\g{\gamma}
\def\l{\lambda}
\def\m{\mu}
\def\n{\nu}
\def\a{\alpha}
\def\eps{\epsilon}
\def\csi{\xi}
\def\gmn{g\_{\mu\nu}}
\def\rar{\rightarrow}
\def\pd#1#2{{\partial#1 \over \partial#2}}
\def\Fi{\FF\_{i}}
\def\ends#1{\(#1\)\_{ends}}
\def\fsi{f\_{i}}
\def\qp{q\_{p}}
\def\ai{\alpha\_{i}}
\def\mi{m\_{i}}
\def\ri{r\_{i}}

\def\cxi{x\_{i}}
\def\ei{e\_{i}}
\def\vri{\vec r\_{i}}
\def\rial{r\_{i}\^{\a}}
\def\etai{\eta\^{(i)}}
\def\Aci{\pd{\A\_{c}\^{i}}{c}\biggr\vert\_{c=1}}
\def\Acis{\pd{\A\_{c}\^{i}}{c}\bigr\vert\_{c=1}}
\def\Ac{\pd{\A\_{c}\^{\a\_{1}...\a\_{d}}}{c}\biggr\vert\_{c=1}}
\def\Acal{\A\_{c}\^{\a\_{1}...\a\_{d}}}
\def\ell{l}
\vsize 23.6truecm
\hsize 5.6truein
\parskip 0pt
\def\oddmargin{.3in}
\def\evenmargin{.1in}

\def\vr{virial relations}
\def\apj{Astrophys. J.}
\def\bk{\par\noindent}
\vskip 0.5truein
\centerline{{\bf  THE VIRIAL THEOREM FOR ACTION-GOVERNED THEORIES}}
\vskip 0.5truein
\centerline{ Mordehai Milgrom
\footnote{*}{Permanent address: Department of condensed-matter physics,
Weizmann Institute, Rehovot 76100 Israel}}
\bk
\centerline{D\'epartment d'Astrophysique Relativiste et de Cosmologie}
\centerline{UPR 176 CNRS, Observatoire de Paris, 92195 Meudon, France}
\vskip 0.5truein
\vskip 30pt
\baselineskip 10pt
\ninerm
{\bf \noindent Abstract.}
We describe a simple derivation of \vr , for arbitrary physical systems
that are governed by an action. The \vt\
may be derived directly from the action, $\A$, with no
need to go via the \eoms , and is simply a statement of the stationarity
of the action
with respect to certain variations in the \dofs. Only
a sub-class of solutions that obey appropriate boundary conditions
satisfy each virial relation.
There is a set of basic \vr\ (one for each
\dof ) of the form $\Acis=0$, where $\A\^{i}\_{c}$ is the action
in which one \dof\ has been scaled by a constant factor $c$.
Linear combinations of these may be put in simple forms, taking advantage
of homogeneity properties of the action, and of dimensional
 considerations.
When some of the \dofs\ are of the same type a tensor \vt\ presents
itself; it may be obtained
in a similar way by considering the variation of the action under linear
transformations among these \dofs . Further generalizations are
 discussed. Symmetries of the action may lead to identities involving
 the \vr . Beside pointing to a unified provenance of the \vr , and
 affording  general systematics of them, our method is a simple
 prescription for deriving such relations. It is particularly useful for
 treating high-derivative and non-local theories.
We bring several examples to show that indeed the usual \vr\ are obtained
by this procedure, and also to produce some new \vr .
\vskip 20pt
{\nineb 1. Introduction}
\vskip 5pt
\par
In default of a general definition of the \vt\
we describe it,
drawing from known examples (see, for some of many,
\REFS{\gold}{H. Goldstein {\it Classical Mechanics}, Addison Wesley,
Singapore (1980).}
\REFSCON{\llft}{L.D. Landau and E.M. Lifshitz {\it The Classical Theory
 of Fields,} (Pergamon, Oxford) (1962).}
\REFSCON{\chafer}{S. Chandrasekhar and E. Fermi, \apj\ {\bf 118} (1953),
 116.}
\REFSCON{\chan}{S. Chandrasekhar, \apj\ {\bf 142} (1965), 1488.}
\REFSCON{\bona}{S. Bonazzola, \apj {\bf 182} (1973), 335.}
\REFSCON{\alm}{M.A. Abramowicz, J.P. Lasota, and B. Muchotrzeb, Commun.
 Math. Phys. {\bf 47} (1976), 109.}
\REFSCON{\vil}{C. Vilain, \apj\ {\bf 227} (1979), 307.}
\REFSCON{\katz}{J. Katz, Physicalia Magazine {\bf 12} (1990), 123
({\it The Gardner of Eden}, Special issue in honor of R. Brout's 60th
birthday, eds. P. Nicoletopoulos, and J. Orloff).}
\REFSCON{\gourg}{E. Gourgoulhon, and S. Bonazzola, Class. and Quant.
Grav. (1994).}\refsend, and, in particular,
\REF{\coll}{G.W. Collins {\it The virial theorem in stellar
 astrophysics,} (Pachart Publication House, Tucson) (1978).}\refend ):
It consists of a set of global (integral) relations that are satisfied
 by a subclass of solutions of the \eoms .
This sub-class is defined by requiring that the solutions obey certain
boundary requirements. The derivation of these relations proceeds as
 follows:
One contracts the \eoms\ with  functions of the \dofs\ and integrate over
the variables on which the latter depend. One then integrates by parts
so as to reduce the order of derivatives,
as many times as possible--discarding boundary terms, as one proceeds,
by imposing requirements on the boundary behavior of the solutions.
One ends up with a virial theorem consisting of (1) the set of integral
relations, and (2) the set of boundary conditions under which they apply.
\par
The applications of this poor-man's substitute for the \eoms\ are many
 and varied (see e.g. \(\chafer\)
\REFS{\chaII}{S. Chandrasekhar, \apj\ {\bf 142} (1965), 1513.}
\REFSCON{\bgsm}{S. Bonazzola, E. Gourgoulhon, M. Salgado, J.A. Marck,
Astron. Astrophys. {\bf 278} (1993), 421.}\refend
\REFSCON{\dyn}{M. Milgrom, Annals of Physics (1994), In press.}\refsend,
and the many uses described in \(\coll \) ).
Its usefulness draws partly from the fact that it involves derivatives
of the \dofs\ of lower order than appear in
the \eoms . (We shall show, in fact, that
the order of derivation appearing is the same as that in the action
 itself.) Thus, for example, the applications in astrophysics require
 knowledge of only
particle velocities, which can be measured, and not of the accelerations,
which cannot. This feature also makes the \vt\ a useful tool for checking
numerical solutions of the \eoms .
\par
Here, we give a systematic treatment of the \vt , for arbitrary systems
governed by equations that are derived
from a variational principle, in a way that highlights
its origin, and greatly facilitate its derivation in the
case of higher order theories, and especially for non-local theories.
\par
We explain how the \vr\ emerge from the action
in sections 2 and 3, showing that they are tantamount to stationarity of
 the action under certain increments in the \dofs ; when the variation of
 the action under such increments may be expressed simply, a convenient
 virial relation results. We demonstrate the procedure with
 various examples in section 4.
\vskip 20pt
{\nineb 2. Derivation of the \vr\ from the action}
\vskip 5pt
\par
Virial relations may be derived for action-governed systems as follows:
The system is described by $d$ \dofs\
$\fsi\(\etai\),~~~1\leq i\leq d$, where, for each $i$,
$\etai$ is the set of variables ,
$\etai\_{1},...$, on which $\fsi$ depends (time, path length,
 space-time coordinates for fields, Fourier variables
if the \dof\ is described in some Fourier space, etc.). One assumes that
there exits an action $\A$: a functional of $\fsi$,
such that under a variation $\d\fsi$ in $\fsi$
$$\d \A=\sum\_{i}\int d\etai~ Q\_{i}~ \d \fsi +
\sum\_{i}\ends{\Fi(\d\fsi)},   \eqn{\i} $$
where $Q\_{i}(f\_{1},...,f\_{d},\etai)$ is a functional of the \dofs\
and a function of $\etai$ only;
$\Fi$ is a linear operator acting on $\d\fsi$, and may depend
as a functional on all the $\fsi$s.
The term $\ends{~~}$ is a boundary term evaluated at the boundary of the
respective $\etai$ space.
One then postulates that the physical states are described by
$\fsi$s for which $\A$ is stationary under variations $\d\fsi$ that
annihilate the ends term. These $\fsi$s are then the solutions of
the \eoms\
$$Q\_{i}(f\_{1},...,f\_{d},\etai)=0~~~~1\leq i\leq d.   \eqn{\ii}  $$
\par
Consider now, as a specific change, a rescaling of a single \dof\ $\fsi$
$$\fsi\rar (1+\eps)\fsi=\fsi+\d\fsi.  \eqn{\iii} $$
The change in the action is then, by eq.\i
$$\d\A=\eps\int d\etai~ Q\_{i}~ \fsi +
\eps\ends{\Fi(\fsi)}.   \eqn{\iv} $$
On the other hand, we can write
$$\d\A=\eps\Aci,  \eqn{\v} $$
where
$$\A\_{c}\^{i}(f\_{1},...,\fsi,...,f\_{d})\equiv
\A(f\_{1},...,c\fsi,...,f\_{d}).  \eqn{\vi} $$
We thus have
$$\Aci=\int d\etai~ Q\_{i}~ \fsi +
\ends{\Fi(\fsi)}.   \eqn{\vii} $$
\par
Equation \vii is but an identity. If we now specialize to solutions of
 the \eoms\ \ii we get
$$\Aci= \ends{\Fi(\fsi)},   \eqn{\viii} $$
which is non-trivial (i.e. is not satisfied by generic
$\fsi\(\etai\)$ that are not solutions of the \eoms\ ).
It tells us that certain expressions derived from the action are pure
boundary terms.
\par
The virial relation  results when we further restrict ourselves to
solutions for which the ends terms may be
discarded. This may happen because the particular $\fsi$s
satisfy boundary conditions in $\etai$ space that render
$\ends{\Fi(\fsi)}$ zero, or, more generally,
because for a large enough $\etai$ volume
the ends term remains finite while $\A$ becomes infinite (in which case
we normalize by the volume, and the ends term vanish in the limit).
We then obtain as our desired \vr\
$$\Aci=0~~~1\leq i\leq d ,  \eqn{\ix}$$
{\it each satisfied on its own sub-class of solutions for which the ends
 term can be discarded.} Clearly, eq.\ix is just
the relation we would get by taking the \eom\ derived by varying $\A$
 with respect to $\fsi$, then multiplying by $\fsi$, then integrating by
 parts as many times as needed to reduce the order of derivatives back to
 that in $\A$. The present method is then but a way to shortcut this
 return journey to the \eoms .
\par
Linear combination of relations\ix with coefficients $\ai$--which should
hold under the appropriate boundary behavior--may be conveniently
written as
$$\Ac=0,  \eqn{\x} $$
where
$$\Acal\equiv\A(c\^{\a\_{1}}f\_{1},...,c\^{\a\_{d}}f\_{d}). \eqn{\xia} $$
Some of these linear combinations may be more useful than others.
For example, some may be reduced to simpler forms by employing
homogeneity properties of the action, or its parts.
If there is a choice of $\ai$ for which
$\Acal=c\^{\b}\A$, with $\b \not= 0$, then, by eq.\x, $\A=0$ is one of
 the \vr . (If $\b=0$ the $d$ relations\ix are dependent, but the theory
 is then degenerate.) This is the case for any system with a Newtonian
 kinetic action and harmonic forces,
in which case $\A=0$ implies the equality of mean kinetic and potential
energies. It is also the case for the Einstein-Hilbert action in vacuum,
in which case the resulting relation is just the integral of the trace
of the \eom .
\par
More generally, the action may be split into terms
$$\A=\sum\_{a}\A\_{a},  \eqn{\xii} $$
all homogeneous with the same set of $\ai$s, but each with its own
 $\b\_{a}$. Then
 $$\sum\_{a}\b\_{a}\A\_{a}=0  \eqn{\xiii} $$
is one of the \vr . This is the case for
a system of particles with a Newtonian kinetic action, and forces that
are homogeneous in the positions (e.g power-law forces). Equation\xiii
 then tell us that the ratio of mean kinetic and potential energies is
 fixed for the theory.
It is also the case for General Relativity with matter in the form of
charged particles, and electromagnetic fields (see sec. 6).
\par
We may also apply dimensional analysis to obtain some
useful combinations of the \vr .
Suppose the action depends on
some constants $\qp,~~ 1\leq p\leq n$. Consider then, for instance,
how the different quantities change under a change in the units
of length by a factor $c$. Depending on their respective dimensions,
$\fsi$ will
change to $c\^{\ai}\fsi$, $\qp$ will change to $c\^{\g\_{i}}\qp$,
while $\A$ will change to  $c\^{\b}\A$. (We assume that $\etai$
have no length dimension, as when they denote the time, or if they do
we redefine them not to have such dimension, and then we have to
add to the $\qp$ another constant of the dimensions of length.)
Dimensional analysis then tells us that
$$\A(c\^{\a\_{1}}f\_{1},...,c\^{\g\_{1}}q\_{1},...)=c\^{\b}\A,
\eqn{\xiv} $$
and so
$$\Acal=c\^{\b}\A(f\_{1},...,f\_{d},c\^{-\g\_{1}}q\_{1},...,c\^{-\g\_{n}}
q\_{n}).  \eqn{\xv}  $$
  From this and eq.\x we then obtain
$$\b\A-\sum\_{p=1}\^{n}\g\_{p}\qp\pd{\A}{\qp}=0,  \eqn{\xvi} $$
as one of the \vr . This is particularly useful when $\A$ depends
only on a small number of constants.
\vskip 20pt
{\nineb 3. Tensor \vt s and further generalizations}
\vskip 5pt
\par
Often, some of the \dofs\ are of the same type--say $\fsi$ for
 $1\leq i\leq K$,
collectively designated $\vec f$. Then, further \vr\ suggest themselves.
(Examples of \dofs\ of the same type are the position coordinates
of a particle, or indeed, of different particles, the components
of the electromagnetic vector potential field, the elements of the
metric tensor etc..)
\par
We now consider the variation of the action under
variations of the form
$$\vec f\rar \M\vec f, \eqn{\xxxi}$$
where $\M=\one+\EPS$ (with $\EPS$ infinitesimal) is a constant
$K\times K$ matrix. From eq.\i we have
$$\d\A=\sum\_{ij}\EPS\_{ij}\left\(\int d\eta~Q\_{i}~f\_{j}
+\ends{\Fi(f\_{j})}\right\).  \eqn{\xxxii} $$
(All the $Q\_{i}$ and $\fsi$, $i\leq K$, are defined on the same $\eta$
space.)

\par
On the other hand, defining
$$\AM(\vec f,f\_{K+1},...,f\_{D})\equiv
\A(\M\vec f,f\_{K+1},...,f\_{D}),  \eqn{\xxxiii} $$
we see that
$$\d\A=\sum\_{ij}\EPS\_{ij}V\_{ij},  \eqn{\xxxiv} $$
where
$$V\_{ij}\equiv\pd{\AM}{\M\_{ij}}\biggr\vert\_{\Msmall\_{ij}
=\d\_{ij}}.  \eqn{\xxxv} $$
Thus
$$V\_{ij}=
\int d\eta~Q\_{i}~f\_{j} +\ends{\Fi(f\_{j})}.  \eqn{\xxxvi} $$
As in section 2, we confine ourselves to solutions of the \eoms\
for which the ends term in eq.\xxxvi can be discarded and then
$$V\_{ij}=0,  \eqn{\xxxvii} $$
which is the desired $K\times K$-tensor \vt (the diagonal elements of
relation\xxxvii are the same as eq.\ix).
In addition we still have, of course, the relations of type \ix for all
the \dofs\ $\fsi~~i>K$.
\par
Symmetries of the action under transformations of $\vec f$ lead to
constraints on the tensor \vr . For example, say that $\A$ is invariant
under a rotations in $\vec f$ space, i.e., under $\vec f\rar U\vec f$,
 for any orthogonal matrix (continuously connected to the unit matrix).
Then, from eq.\xxxiv, $\sum\_{ij}\EPS\_{ij}V\_{ij}=0$ for any
antisymmetric matrix $\EPS$ (the generators of the orthogonal
 transformations). Thus, $V$ must be a symmetric matrix:
 $V\_{ij}=V\_{ji}$ is then an identity,
and there are only $K(K+1)/2$ independent relations.
Note that symmetries of the action do not necessarily imply symmetries
of the above type where only some of the \dofs\ are transformed.
\par
The information held by the \eoms\ on their solutions is infinitely
 greater than that supplied by the finite number of discrete, global
 relations that we have discussed so far. It should be obvious then
 that one can derive an infinite number of further such relations.
These emerge if we consider the variation of the action under
general infinitesimal increments of the form
$$\fsi\rar\fsi+\csi\_{i}(f\_{1},...,f\_{d}).  \eqn{\xxxviii} $$
Here $\csi\_{i}$ may be a function of the \dofs\ and also a functional of
them, but it must, of course, be ``of the same type'' as $\fsi$ itself
 (in the sense that the action is defined if we replace $\fsi$ by
 $\csi\_{i}$). As before, if the ends term $\ends{\Fi(\csi\_{i})}$ may be
 discarded, the variation of $\A$ under \xxxviii vanishes on solutions
 of the  \eoms .
When, for given $\A$ and $\csi$,  $\d\A$ can be written in a
simple form--as was the case for the linear increments we had discussed
earlier--a useful relation may be obtained. The point is that because the
increment \xxxviii is defined in terms of the \dofs\ themselves it
 enables us to formulate the extremum condition on $\A$, and the
 required boundary conditions, in terms of the $\fsi$s themselves thus
 leading to a virial-like relation.
When it is not necessary to have a relations written in terms of system
properties alone, it may be useful to consider more general increments
in which $\csi\_{i}$ is also an explicit function of $\etai$.
Thus considering $\fsi\rar \eps\_{ijk}f\_{j}f\_{k}$
will lead to higher-tensorial-order relations (for a system of Newtonian
particles they correspond to relations connected with higher moments of
the mass distribution--see section 5).
\par
While these further relations are not as amenable to many of the uses as
their predecessors, they may be useful, depending on the particular case.
For example, in connection
with checks on numerical solutions of the \eoms\ the may provide
checks of aspects not covered by the basic relations (see section 4).
\par
We have discussed \vr\ as they apply to the full sub-class of solutions
that satisfy the appropriate boundary conditions.
When one limits himself further to special solution--such as ones with
certain symmetry properties--the \vt\ may, in general, be simplified;
we shall not deal with such further restrictions.
\vskip 20pt
{\nineb 4. Examples}
\vskip 5pt
\par
Now let us look at some examples, mainly
to demonstrate the procedure and various relevant points discussed
generally in the previous sections.
We start with the archetype of all \vt s: that for a system of $N$
Newtonian particles interacting via a general $N$-body potential.
The action for a finite time span $(t\_{1},t\_{2})$ is
$$\A=\A\_{K}+\A\_{P},  \eqn{\xli} $$
with the kinetic action
$$\A\_{K}=T\^{-1}\int\_{t\_{1}}\^{t\_{2}}\oot \sum\_{i}\mi\left(
\deriv{\vri}{t}\right)^2~dt,  \eqn{\xlia} $$
and the potential action
$$\A\_{P}=-T\^{-1}\int\Phi\(\vrt{1},...,\vrt{N}\)~dt.     \eqn{\xlib} $$
We have normalized the standard action by $T\equiv t\_{2}-t\_{1}$,
to make $\A$ finite for $T\rar\infty$.
\par
The \dofs\ are the $3N$ components $\rial$ of the positions.
We get, straightforwardly, from eq.\ix the $3N$ \vr\
$$\av{\mi(\dot\rial)\^{2}}-\av{{\rial\pdline{\Phi}{\rial}}}=0,
\eqn{\xlii} $$
where
$$\av{Q}\equiv T\^{-1}\int\_{t\_{1}}\^{t\_{2}}Q~dt.  \eqn{\xliia} $$
Usually this average is taken in the limit $t\_{1}\rar -\infty$, and
$t\_{2}\rar\infty$, and it then holds for solutions in which all
 particles are bounded in a finite volume.
The standard scalar \vt\ for this case:
$$\avlar{\sum\_{i}\mi v\_{i}\^{2}}-\avlar{\sum\_{i}\vri\cdot\grad\_{\vri}
\Phi}=0,  \eqn{\xliii} $$
is obtained by differentiating $\A\(c\vrt{1},...,c\vrt{N}\)$ with respect
 to $c$ (at $c=1$). There is a $3N\times 3N$ tensor \vt\
$$\av{\mi \dot\rial\dot r\_{j}\^{\b}}-\av{\rial\pdline{\Phi}{r\_{j}\^
{\b}}}=0, \eqn{\xliv} $$
which is obtained from eq.\xxxvii, by
considering $\rial\rar\rial +\eps r\_{j}\^{\b}$,
for fixed values of all indices.
Neither term is symmetric when the masses are not all equal, because
neither term in the action is symmetric under rotations in configuration
space. The standard tensor \vt\ is obtained by taking the trace over
the particle indices, or by considering $\rial\rar\rial+\eps
 r\_{i}\^{\b}$ for all values of i but fixed $\a,\b$.
$$\avlar{\sum\_{i}\mi \dot\rial\dot r\_{i}\^{\b}}-\avlar{\rial
\pdline{\Phi}{r\_{i}\^{\b}}}=0.  \eqn{\xlivb} $$
The kinetic term is now identically symmetric because the kinetic action
is symmetric under simultaneous rotations of all the $\vec r\_{i}$.
The antisymmetric part of the potential term is
$\sum\_{i}\vec r\_{i}\times\grad\_{\vec r\_{i}}\Phi$ which vanishes
identically if $\Phi$ is rotationally invariant, in which case relation
\xlivb is identically symmetric in $\a\b$.
\par
We now add a field \dof\ by considering a system of $N$ gravitation
particles of masses $\mi$ and we also want to solve for the (Newtonian)
gravitational field $\varphi(\vec r)$ which is thus also a \dof .
The action is
$$\A=T\^{-1}\int dt\left\{\sum\_{i}\oot\mi v\_{i}\^{2}
-\sum\_{i}\mi\varphi(\ri)-{1\over 8\pi G}\int(\grad\varphi)\^{2}
\dtr\right\}. \eqn{\xlv} $$
It gives the standard \eom\ of the particles in the potential $\varphi$,
and variation over $\varphi$ gives the Poisson equation for $\varphi$
with the density $\rho(\vec r)=\sum\_{i}\mi\d\^{3}(\vec r-\vec r\_{i})$.
\par
Beside the scalar and tensor \vr\ we wrote in the previous example
(with $\Phi\(\vrt{1},...,\vrt{N}\)=\sum\_{i}\mi\varphi\(\vrt{i}\)$),
we now have a relation resulting from $\varphi\rar c\varphi$
which is read directly from the action--the three terms in $\A$ being
homogeneous in $\varphi$, with powers 0,2, and 1 respectively:
$$\avlar{\sum\_{i}\mi\varphi(\ri)}+{1\over 4\pi G}
\avlar{\int(\grad\varphi)\^{2}\dtr}=0. \eqn{\xlvi} $$
To demonstrate the potential usefulness of relations resulting from the
more general types of variations eq.\xxxviii, consider an infinitesimal
increment
$$\varphi\rar\eps J(\varphi). \eqn{\xlvii} $$
Substituting in the action, and taking the first order in $\eps$
we get
$$\avlar{\sum\_{i}\mi J\(\varphi(\ri)\)}+{1\over 4\pi G}
\avlar{\int J'(\varphi)(\grad\varphi)\^{2}\dtr}=0, \eqn{\xlviii} $$
with the boundary requirement being that $J(\varphi)=0$.
Clearly, this infinite set of relations captures more of the information
in the \eom . By choosing $J$ properly we may accentuate different
regions of space, for example regions where there are better
 measurements, or where better approximations can be made.
\par
Consider now a system of particles governed by a general kinetic action,
so the action is of the form
$$\A=\A\_{K}\(\vrt{1},...,\vrt{N},q\_{1},...,q\_{n}\)
-T\^{-1}\int~\Phi\(\vrt{1},...,\vrt{N}\)~dt.   \eqn{\txxx} $$
where $\A\_{K}$ is a general functional of the particle trajectories
$\vrt{i}$ (including higher-derivative or, indeed, non-local actions,
in which case the full world lines of the particles are integrated on),
and $\qp$ are constants.
We apply eq.\xvi  noting that $\b=2$ (with our choise of normalization
$\A$ has dimensions of energy), and that $\Phi$ must be of the form
$$\Phi=\Phi\_{0}\phi(\vec r\_{1}/\ell,...,\vec r\_{N}/\ell).
 \eqn{\txxxi} $$
Here, $\Phi\_{0}$, and $\ell$ are constants with the dimensions of energy
and length, respectively.
Thus, we obtain
in generalization of the scalar \vt\
resulting from $\vrt{i}\rar(1+\eps)\vrt{i}$
 (see \(\dyn \))
$$2\A\_{K}-\sum\_{p=1}\^{n}\g\_{p}\qp\pd{\A\_{K}}{\qp}-\avlar{\sum
\_{i=1}\^{N}\vec r\_{i}\cdot\grad\_{\vec r\_{i}}\Phi}=0. \eqn{\txxxii} $$
As in eq.\xvi , $\g\_{p}$ are the powers of length in the dimensions of
 the constants $\qp$.
\par
When $\A\_{K}$ is a Lagrangian action:
$$\A\_{K}=T\^{-1}\int L\_{K}\(\vec r\_{i},\vec r\_{i}\^{(1)},
\vec r\_{i}\^{(2)},...,\qp\)~dt,  \eqn{\txxxiii}$$
with the kinetic Lagrangian $L\_{K}$ depending on higher time derivatives
$\vec r\_{i}\^{(\ell)}$ of the trajectories, we can write,
in generalization of the standard scalar \vt , either
$$\avlar{\sum\_{i=1}\^{N}\sum\_{\ell}\vec r\_{i}\^{(\ell)}\cdot
\pd{L\_{K}}{\vec r\_{i}\^{(\ell)}}}-\avlar{\sum\_{i=1}\^{N}
\vec r\_{i}\cdot\grad\_{\vec r\_{i}}\Phi}=0, \eqn{\txxxiv} $$
obtained from the variations of $\A$ under
$\vrt{i}\rar c\vrt{i},~~1\le i\le N$,
or, equivalently, from eq.\txxxii ,
$$2\av{L\_{K}}-\sum\_{p=1}\^{n}\g\_{p}q\_{p}\pd{\av{L\_{K}}}{q\_{p}}
-\avlar{\sum\_{i=1}\^{N}
\vec r\_{i}\cdot\grad\_{\vec r\_{i}}\Phi}=0. \eqn{\txxxv} $$
The latter is particularly useful when $L\_{K}$ depends only on a small
number of constants.
\par
The tensor \vt , also read directly from $\A$ is
$$V\_{\a\b}\equiv\avlar{\sum\_{i=1}\^{N}\sum\_{\ell}r\_{i}\^{\a(\ell)}
\pd{L\_{K}}{r\_{i}\^{\b(\ell)}}}-\avlar{\sum\_{i=1}\^{N}
r\_{i}\^{\a}\pd{\Phi}{r\_{i}\^{\b}}}=0 \eqn{\txxxvi} $$
($r\_{i}\^{\a (\ell)}$ is the $\ell$th time derivative of the $\a$
 component of $\vec r\_{i}$).
Rotational invariance of $\A\_{K}$ implies that the kinetic part of
$V\_{\a\b}$ is symmetric, as an identity, although this is not manifest
 in eq.\txxxvi .
\par
For a general relativistic system of particles of masses $\mi$
and electric charges $\ei$,
the action is\REF{\wein}{S. Weinberg {\it Gravitation an Cosmology,}
(John Wiley, New York) (1972).}\refend
$$\A=\A\_{G}+\A\_{p}+\A\_{EM}+\A\_{e},   \eqn{\vbi} $$
where
$$\A\_{G}=-{1\over 16\pi G}\int R\dfxg   \eqn{\vbii} $$
is the Einstein action for the metric;
$$\A\_{p}=-\sum\_{i}\mi\int d\l\left\(-\gmn(x\_{i})
\deriv{\cxi\^{\m}}{\l}\deriv{\cxi\^{\n}}{\l}\right\)^{1/2} \eqn{\vbiii}$$
is the particle action;
$$\A\_{EM}=-{1\over 4}\int (A\_{\n ,\m}-A\_{\m ,\n})g\^{\m\a}g\^{\n\b}
(A\_{\b ,\a}-A\_{\a ,\b})\dfxg   \eqn{\vbiv}  $$
is the action for the electromagnetic field; and
$$\A\_{e}=\sum\_{i}\ei\int d\l\deriv{\cxi\^{\m}}{\l}A\_{\m}(\cxi)
 \eqn{\vbv}$$
is the field-particle-interaction contribution.
The \dofs\ here are the particle space-time coordinates $\cxi\^{\m}(\l)$,
the four electromagnetic-potential fields $A\_{\m}(x)$, and the
components, $\gmn(x)$, of the metric tensor.
\par
Under $\gmn\rar c\gmn$ the terms in the action are homogeneous:
$\A\_{G}\rar c\A\_{G},~~\A\_{p}\rar c\^{1/2}\A\_{p}$, and $\A\_{EM}$, and
$\A\_{e}$ are invariant. Thus we get as one of the \vr\
$$\A\_{G}+\oot\A\_{p}=0.  \eqn{\vbvi} $$
This can be shown to be just the space-time integral of the trace of
the (Einstein) \eom . This result is easily seen to be general: The
 definition
of the energy-momentum tensor, $T\_{\m\n}$, is such that under a change
$\d\gmn$ we have $\d\A\equiv (1/2)\int\d\gmn T\^{\m\n}\dfxg$ \(\wein\) ;
so, under $\gmn\rar (1+\eps)\gmn$, $\d\A=(\eps/2)\int T\dfxg$, for
 whatever the source is. The boundary requirements for this type of
 relation is satisfied if the metric becomes flat at spatial infinity (so
 no  gravitational waves escape to infinity), and if there are periodic
 boundary conditions in time (e.g. a stationary solution).
\par
There is also homogeneity under $A\_{\a}\rar cA\_{\a}$, which gives a
virial relation
$$2\A\_{EM}+\A\_{e}=0  \eqn{\vbvii} $$
(requiring that no EM waves escape to infinity).
\par
There are tensor \vt s for $\gmn$ and for $A\_{\a}$; the latter, for
 example, gotten by considering $A\_{\a}\rar A\_{\a}+\eps A\_{\b}$ reads
$$V\_{\b}\^{\a}\equiv -\int A\_{\b ,\m}g\^{\m\g}g\^{\a\n}(A\_{\n ,\g}
-A\_{\g ,\n})\dfxg  +
\sum\_{i}\ei\int d\l\deriv{\cxi\^{\a}}{\l}A\_{\b}(\cxi)=0.\eqn{\vbviii}$$
All these relations may be easily derived from the \eoms , as is always
the case with low-order theories. We bring them here to show how they
follow from our systematic procedure, by mere inspection of the action,
and without having to derive the \eoms\ first.
\par
By employing a more general conformal increment
$$\gmn\rar\gmn+\eps J(\gmn,A\_{\a},..)\gmn,  \eqn{\vbviit} $$
eq.\vbvi generalizes to
$$\int\(-(8\pi G)\^{-1}R+T\)J\dfxg=0, \eqn{\vbix} $$
and there is a similar generalization for eq.\vbvii .
\par
The action may have terms quadratic in the curvature tensor--as is the
case for the first order semi-classical corrections to General Relativity
obtained from string theory--such as
$$\int R\^{2}\dfxg,~~~\int R\_{\m\n}R\^{\m\n}\dfxg,~~~
\int R\_{\a\b\m\n}R\^{\a\b\m\n}\dfxg.  \eqn{\vbvia} $$
These are invariant under $\gmn\rar c\gmn$, and
do not modify relation \vbvi . In order to capture them
in a virial relation we may need to employ a tensor relation, or use
one that is induced by increments of type \xxxviii .
\vskip 5pt
{\bf Acknowledgement}
I thank Brandon Carter, Eric Gourgoulhon, Jean-Pierre Lasota, and
Thibault Damour for useful discussions, and Jacob Bekenstein for
a critical reading of the manuscript. The hospitality
of the department of relativistic astrophysics and cosmology
at the Paris Observatory is gratefully acknowledged.

\endpage
\endpage
\write2{\bigskip\noindent
  {\tofcfont References\leadtofc\number\count0}\par\smallskip}
\chskipt
\ifodd0{
\rjustline{{\chapfont References}}}
\else{
\ljustline{{\chapfont References}}}\fi
\chskipl
\immediate\closeout 4
\input refc
\endpage
\bye